\documentclass{article}
\usepackage{spconf,amsmath,graphicx,hyperref}
\usepackage{amssymb}
\usepackage{algorithm}
\usepackage{algpseudocode}
\usepackage{graphicx}
\usepackage{subcaption}
\usepackage{xcolor}

\newcommand{\qa}{{\bf a}}

\newcommand{\qc}{{\bf c}}

\newcommand{\qg}{{\bf g}}
\newcommand{\qh}{{\bf h}}

\newcommand{\qv}{{\bf v}}
\newcommand{\qw}{{\bf w}}
\newcommand{\qx}{{\bf x}}

\newcommand{\qG}{{\bf G}}
\newcommand{\qH}{{\bf H}}

\newcommand{\qR}{{\bf R}}

\newcommand{\twsj}{{\Tilde{\qw}_{sj}}}
\newcommand{\twspj}{{\Tilde{\qw}_{s'j}}}

\newcommand{\bas}{\bar{a}_s}
\newcommand{\baps}{\bar{a}_{s'}}
\newcommand{\tas}{\Tilde{a}_s}
\newcommand{\taps}{\Tilde{a}_{s'}}
\newcommand{\bgsk}{\bar{\qg}_{sk}}
\newcommand{\bgspk}{\bar{\qg}_{s'k}}
\newcommand{\tgsk}{\Tilde{\qg}_{sk}}

\newcommand{\bwsi}{{\bar{\qw}_{si}}}

\newcommand{\bwsk}{{\bar{\qw}_{sk}}}

\newcommand{\rhokjs}{\rho_{kj}^{ss}}
\newcommand{\rhokjssp}{\rho_{kj}^{ss'}}

\newcommand{\veks}{\varepsilon_k^s}

\newcommand{\tkis}{t_{ki}^s}
\newcommand{\vrhokjssp}{\varrho_{kj}^{ss'}}
\newcommand{\vrhokjs}{\varrho_{kj}^{ss}}

\newcommand{\betasi}{\bar{\eta}_{si}}
\newcommand{\betask}{\bar{\eta}_{sk}}

\newcommand{\Tetasj}{\Tilde{\eta}_{sj}}
\newcommand{\betak}{\bar{\eta}_{k}}
\newcommand{\betai}{\bar{\eta}_{i}}

\newcommand{\Tetaspj}{\Tilde{\eta}_{s'j}}
\newcommand{\Set}{\mathcal{S}}
\newcommand{\Kn}{\mathcal{K_N}}
\newcommand{\Kf}{\mathcal{K_F}}

\newcommand{\Ex}{\mathbb{E}}

\newcommand{\ZF}{\mathtt{ZF}}
\newcommand{\tr}{\mathtt{tr}}

\newcommand{\Tetak}{\Tilde{\eta}_k}
\newcommand{\Tetaj}{\Tilde{\eta}_j}

\newcommand{\tSEkth}{\tilde{\mathcal{R}}_{k}}
\newcommand{\bSEkth}{\bar{\mathcal{R}}_{k}}

\title{Power consumption Reduction in ELAA-Assisted ISAC Systems }

\name{Xiaomin Cao, Mohammadali Mohammadi, Hien Quoc Ngo, and Michail Matthaiou \thanks{This work was supported by the European Research Council (ERC) under the European Unions Horizon 2020 research and innovation programme (grant agreement No. 101001331).}}
\address{Centre for Wireless Innovation (CWI), Queen's University Belfast, U.K.}

\begin{document}

\maketitle
\begin{abstract}

In this paper, we consider power consumption reduction in extremely large antenna arrays (ELAAs) for integrated sensing and communication (ISAC) applications. Although ELAAs are critical for achieving high-resolution near-field sensing, fully activating all antenna elements in conventional digital architectures leads to prohibitive power demands. To address this, we propose an energy-efficient subarray activation framework that selects an optimal subset of subarrays to minimize the total power consumption, subject to quality-of-service (QoS) constraints for both sensing and communication. We formulate a novel optimization problem and solve it using a successive convex approximation (SCA)-based iterative algorithm. The simulation results confirm that the proposed method significantly reduces power consumption while maintaining dual-function performance.

\end{abstract}
\vspace{-0.5em}
%
\vspace{-0.5em}
\section{Introduction}
\label{sec:intro}
\vspace{-0.5em}

ELAAs are central to realizing near-field ISAC, where spherical wavefronts introduce an additional degree of freedom in the distance domain. This enables spatial information to be encoded for high-precision sensing and localization. Their capability to generate highly focused beams further enhances both communication efficiency and target detection~\cite{Han:IoT:2023, Cong:WCOML:2024}. These advantages position ELAAs as a strong candidate for ISAC, which is expected to serve as a cornerstone of sixth-generation (6G) networks~\cite{Zhang:J_STSP:2021, Fan:JSAC:2022}. Yet, their deployment faces two critical challenges: hardware cost/complexity and high 
power consumption if employing a conventional fully-digital beamforming architecture, where each antenna element requires its own radio frequency (RF) chain.

To mitigate hardware limitations, while maintaining satisfactory performance, hybrid beamforming architectures~\cite{Sohrabi:JSTSP:2016,Gao:JSAC:2016,Liu:JSAC:2025} and  antenna selection techniques~\cite{Wang:J_STSP:2024, Mari:TVT:2020} have been proposed. The two widely studied hybrid architectures, namely fully-connected (where each RF chain is linked to all antennas~\cite{Sohrabi:JSTSP:2016}) and fixed-subarray (where each RF chain connects to a subset of antennas~\cite{Gao:JSAC:2016}), are, however, vulnerable to near-field effects. Specifically, they can only adjust the signal phases while keeping the amplitudes fixed, which limits their adaptability in near-field channels, where the channel gains vary significantly across antennas~\cite{Liu:JSAC:2025}. A fundamental flaw of hybrid beamforming is that its frequency-independent analog phase-shifter network cannot address the frequency-varying phase shifts across the near-field wideband channel~\cite{Luo:TWC:2024}. Moreover, larger arrays make antenna selection more complex, adding significant processing and control overhead.



Considering spatial non-stationarity in ELAAs, where users interact with only part of the array, termed as their visibility region (VR)~\cite{Zhang:TCOM:2024}, partial subarray activation can effectively support users and reduce the overall power consumption~\cite{Mumtaz:WCL:2025}.
Motivated by this, we investigate subarray activation in ELAA-assisted ISAC systems, where an ELAA with multiple fixed-subarrays simultaneously serves near-field (NFUEs) and far-field users (FFUEs) in the downlink, while also illuminating a target for sensing. The main contributions of this work are two-fold: 1) We formulate an optimization problem that aims to minimize the total power consumption by selecting an optimal subset of the subarrays, while ensuring the QoS requirements for both multi-user communication and near-field sensing. To tackle the resulting non-convex and combinatorial problem, we develop an efficient algorithm that delivers a high-quality suboptimal solution with polynomial computational complexity. 2) Our simulation results demonstrate the effectiveness of the proposed scheme, showing that it achieves significant power savings compared to the fully activated ELAA baseline while still meeting the stringent performance requirements of both communication and sensing tasks.

\textit{Notation:} We use bold lowercase (uppercase) letters to denote vectors (matrices). The Hermitian and trace operators are denoted by $(\cdot)^H$ and $\operatorname{tr}(\cdot)$, respectively. A circular, zero-mean, symmetric complex Gaussian distribution with variance $\sigma^2$ is written as $\mathcal{CN}(0,\sigma^2)$. Finally, $\mathbb{E}\{\cdot\}$ denotes statistical expectation.  
\vspace{-0.5em}
\section{System Model}
\label{sec:system model}
\vspace{-0.5em}
We consider a monostatic ELAA-assisted ISAC system designed for passive target sensing, where the base station (BS) supports both downlink communication and sensing functionalities. We consider $K \!=\! K_N + K_F$ single-antenna communication users and one single target, where $K_N$ NFUEs are located in the near-field region of the ELAA, while $K_F$ FFUEs are located in the far-field region of the ELAA. 
The number of transmit antennas is $M_t$, which are divided into $S$ subarrays, each containing $M_s$ antenna elements. For simplicity of notation, we define the sets $\Kn\triangleq\{1,\ldots,K_N\}$, $\Kf\triangleq\{1,\ldots,K_F\}$, and $\mathcal{S}\triangleq\{1,\ldots,S\}$ to represent the sets of NFUEs, FFUEs, and subarrays, respectively.

The joint sensing and communication signal of the $s$-th subarray transmitted at time $l$ is given by \cite{Xiang:TSP:2020}
\vspace{-0.3em}
\begin{align} \label{eq:Tx signal}
    \qx_{l,s} &\!= \sum\nolimits_{k\in\Kn}\!\sqrt{\betask} \bas \bwsk \bar{c}_{k,l} \!+\! \sum\nolimits_{j\in\Kf}\! \sqrt{\Tetasj} \tas \twsj \tilde{c}_{j,l} \nonumber \\
    &\hspace{2em}+  \sqrt{\eta_s} \qw_{s}^r a_s r_{l,s},
\end{align}
where the binary variables $\bas, \tas\in\{0,1\}$ indicate whether the subarray $s$ is being activated for serving the NFUEs and FFUEs, receptively. We assume that the activated communication subarrays are also used for sensing. Accordingly, we set $a_s = \min\{1,\bas+\tas\}$. Moreover, in~\eqref{eq:Tx signal}, $\bwsk$, $\twsj$ and $\qw_{s}^r$ denote the precoding vector at the $s$-th subarray corresponding to the $k$-th NFUE, the $j$-th FFUE and the target, respectively;  $\bar{c}_{k,l}$ and $\tilde{c}_{j,l}$ are the information symbols intended for NFUE $k$ and FFUE $j$ at time instant $l$, which are mutually independent with zero mean and unit variance; $r_{l,s}$ is the sensing symbol transmitted by the $s$-th subarray at time instant $l$; $\betask$, $\Tetasj$ are the power allocation coefficients for NFUE $k$  and FFUE $j$ at sub-array $s$, respectively. 

The near-field channel between the $(m_x, m_y)$-th element of the ELAA and $k$-th NFUE, is given by \cite{Jun:TCom:2024}
\vspace{-0.7em}
\begin{align*}
    \bar{g}_{m_x,m_y}(\theta_k,\phi_k) \!=\! \frac{\lambda}{4\pi r_{k,m_x,m_y}(\theta_k,\phi_k)} e^{-j \frac{2\pi}{\lambda} r_{k,m_x,m_y}(\theta_k,\phi_k)},
\end{align*}
where $\lambda$ is the carrier frequency, $\theta_k$ and $\phi_k$ are elevation and azimuth angles corresponding to the $k$-th NFUE, and
\vspace{-0.5em}
\begin{align}~\label{eq:rkm}
    r_{k,m_x,m_y} \!\!=\!\!\sqrt{\!r_k^2 \!\!-\!\! 2r_k m_x d\cos\theta_k \sin\phi_k \! - \! 2r_k m_y d \sin\theta_k \! + \! \omega},
\end{align}
denotes the distance between the center of $(m_x,m_y)$-th element to the $k$-th NFUE, while $\omega\triangleq(m_x^2 \!+\! m_y^2) d^2$. Moreover,  we denote the overall channel between the ELAA and $k$-th NFUE as $\bar{\qg}_k = [\bar{g}_{1,1}(\theta_k, \phi_k),\ldots,
\bar{g}_{M_x,1}(\theta_k, \phi_k),\ldots, \\ \bar{g}_{M_x,M_y}(\theta_k, \phi_k)]^T$, while $\bar{\qH}_1 \!=\! [\bar{\qg}_1,\ldots,\bar{\qg}_{K_N}] \in \mathbb{C}^{M_t \times K_N}$ represents the aggregate channel of all NFUEs.

The channel vector corresponding to the $k$-th FFUE is denoted by $\tilde{\qg}_{k} = \beta_k^{1/2} \qh_k =[\tilde{\qg}_{1k}^T, \ldots, \tilde{\qg}_{Sk}^T]^T$, where $\beta_k$ denotes the large-scale fading, $\qh_k$ represents the small-scale fading channel with $\mathcal{CN}(0,1)$ distributed elements and $\tilde{\qg}_{sk}\in\mathbb{C}^{M_s\times 1}$ represents the channel between subarray $s$ and FFUE $k$. We denote the overall channel for all FFUEs as $\Tilde{\qG} = [\qh_1, \ldots, \qh_{K_F}]\in\mathbb{C}^{M_t \times  K_F}$.

The received signal at NFUE $k$ can be written as
\vspace{-0.4em}
\begin{align} \label{eq:receive signa:Near}
    \bar{y}_{k,l}  
    &= \sum\nolimits_{s\in\Set}\!   
    \Big(\! \sum\nolimits_{k\in\Kn}\!\!\!\sqrt{\betask}\bgsk^H\bas\bar{\qw}_{sk} \bar{c}_{k,l} \nonumber \\   
    &\hspace{2em} +\!\sum\nolimits_{j\in\Kf}\!\!\!\sqrt{\Tetasj}\bgsk^H\tas\Tilde{\qw}_{sj} \tilde{c}_{j,l}\Big)
    \! \nonumber \\   
    &\hspace{2em} + \sum\nolimits_{s\in\Set} \!\! \sqrt{\eta_s} \bgsk^H \qw_{s} a_s r_{l,s}  + \bar{n}_{k,l}, 
\end{align}
where $\bar{n}_{k,l}\sim\mathcal{CN}(0,\sigma_k^2)$ denotes the additive white Gaussian noise (AWGN). 
The received signal at FFUE $k$ is
\vspace{-0.5em}
\begin{align} \label{eq:receive signa:Far}
    \Tilde{y}_{k,l}   
    &= \sum\nolimits_{s\in\Set}\!     \Big(\!\sum\nolimits_{k\in\Kf}\!\!\sqrt{\Tilde{\eta}_{sk}}\tgsk^H\tas\Tilde{\qw}_{sk}\Tilde{c}_{k,l} \nonumber \\   
    &\hspace{2em}+ \!\sum\nolimits_{i\in\Kn}\!\!\sqrt{\betasi}\tgsk^H
    \bas \bwsi \bar{c}_{i,l}\Big) \nonumber \\   
    &\hspace{2em} +  \sum\nolimits_{s\in\Set} \!\! \sqrt{\eta_s} \tgsk^H \qw_{s} a_s r_{l,s}  + \Tilde{n}_{k,l}, 
\end{align}
where $\Tilde{n}_{k,l}\sim\mathcal{CN}(0,\sigma_k^2)$ denotes the AWGN.

The transmit steering vector pointing towards the target, is given by 
\begin{align}
   &\qv(\theta_s^t, \phi_s^t) =\Big[\frac{\lambda}{4\pi r_{s,M_x,M_y}}e^{-j \frac{2\pi}{\lambda} r_{s,1,1}(\theta_s^t,\phi_s^t)}, \nonumber\\
   &\hspace{2em}\ldots, 
 \frac{\lambda}{4\pi r_{s,M_x,M_y}} 
 e^{-j \frac{2\pi}{\lambda} r_{s,M_x,M_y}(\theta_s^t,\phi_s^t)}\Big]^T, 
\end{align}
where $r_{s,m_x,m_y}$ is obtained from~\eqref{eq:rkm} by replacing $(r_k, \theta_k,\phi_k)$ with $(r_s, \theta_s,\phi_s)$, where $r_s$ is the distance from the target to the array, which is assumed known since we focus on the estimation of azimuth angle $\theta_s^t$ and elevation angle $\phi_s^t$ of the target.

\vspace{-0.5em}
\section{Performance Analysis and Problem Formulation}
\label{sec:performance}
\vspace{-0.5em}
In this section, we first analyze the signal-to-interference-plus-noise ratio (SINR) for communication and beampattern gain for sensing. Then, we formulate an optimization problem aimed at minimizing overall power consumption through intelligent subarray activation, while ensuring a satisfactory performance for both communication and sensing tasks.
\subsection{Performance Analysis}
\label{subsec:SINR}
\vspace{-0.5em}
We apply zero-forcing (ZF) precoding at the BS to design communication beamforming vectors for NFUEs and FFUEs. Specifically, we set the precoding vector of the subarray $s$ for NFUE $k$ as $\bar{\qw}_{sk}^{\ZF} = \big[\bar{\qH}_1(\bar{\qH}_1^{H}\bar{\qH}_1)^{-1}\big]_{(\iota_s,k)} \in \mathbb{C}^{M_s \times 1}$, where $\iota_s$ is the antenna index set corresponding to the $s$-th subarray, defined as $\iota_s \triangleq \{(s-1) M_s+1, sM_s\}$. For FFUEs, we set the precoding vector of the subarray $s$ towards FFUE $k$ as $\Tilde{\qw}_{sk}^{\ZF} = \big[\Tilde{\qG} (\Tilde{\qG}^H\Tilde{\qG})^{-1}\big]_{(\iota_s,k)} \in \mathbb{C}^{M_s \times 1}$.


Note that the VR-based ZF scheme bears resemblance to centralized cell-free massive MIMO systems. Therefore, to eliminate intra-group interference, we set the power control coefficients to $\betak\triangleq\bar{\eta}_{1k}^{\ZF}=\ldots=\bar{\eta}_{Sk}^{\ZF}$ for any NFUE $k$ \cite{2024:Mohammadi:survey}. For FFUEs, $\Tetak\triangleq\Tilde{\eta}_{1k}^{\ZF}=\ldots=\Tilde{\eta}_{Sk}^{\ZF}$ follows similarly.

Accordingly, by using the ZF precoders in~\eqref{eq:receive signa:Near}, the received SINR of the $k$-th NFUE can be written as
\vspace{-0.2em}
\begin{align} \label{eq:SINR:NF:CZF}
   &\overline{\text{SINR}}_{k}= \frac{\betak\big\vert\sum\nolimits_{s\in\Set}\bar{\qg}_{sk}^{H}\bas\bar{\qw}_{sk}^{\ZF}\big\vert^2}{\bar{\Delta}^{\ZF}+ \sigma_k^2}, 
\end{align}
with $\bar{\qg}_{sk}\!=\![\bar{\qg}]_{\iota_s}\!\!\in\! \mathbb{C}^{M_s \times 1}$, $\bar{\Delta}^{\ZF}\!\!\triangleq \!\sum\nolimits_{i \in\Kn\setminus k} \!\! \betai\!\sum\nolimits_{s\in\Set}\! \!
\sum\nolimits_{s'\in\Set}\! \\
\bas \baps\bgsk^{H}  \bar{\qw}_{si}^{\ZF} (\bar{\qw}_{s'i}^H)^{\ZF} \bgspk^{H} 
\!+ \sum\nolimits_{j\in\Kf}\! \Tetaj\! \sum\nolimits_{s \in \Set}\!\!\sum\nolimits_{s' \in \Set}\! \tas \taps \rhokjssp 
+\! \sum\nolimits_{s\in\Set}\! \eta_s a_s^2 \vert\bar{\qg}_{sk}^{H} \qw_s^r\vert^2$, 
while  $\rhokjssp\! \triangleq\! 
\bar{\qg}_{sk}^{H}\Ex\big\{\Tilde{\qw}_{sj}^{\ZF} (\Tilde{\qw}_{s'j}^{\ZF})^H \big\}\bar{\qg}_{s'k} $.

The received SINR at the $k$-th FFUE can be obtained as 
\vspace{-0.8em}
\begin{align} \label{eq:SINR:FF:CZF}
    \widetilde{\text{SINR}}_{k} = \frac{\Tetak\big\vert\Ex\left\{\sum\nolimits_{s\in\Set}\Tilde{\qg}_{sk}^H\tas\Tilde{\qw}_{sk}^{\ZF}\right\}\big\vert^2}{ \tilde{\Delta}^{\ZF}+ \sigma_k^2},
\end{align}
where $\tilde{\Delta}^{\ZF} \!\triangleq\!  \Tetak \!\sum\nolimits_{s\in \Set} \tas \veks + \! \sum\nolimits_{i\in\Kn} \! \betai \! \sum\nolimits_{s\in \Set} \bas^2 \tkis +\! \sum\nolimits_{j\in\Kf\setminus k}\Tetaj \!\sum\nolimits_{s \in \Set}\!\sum\nolimits_{s' \in \Set}\! \tas \taps \vrhokjssp +\!  \sum\nolimits_{s\in \Set} \! \eta_s a_s^2 \beta_k (\qw_s^r)^H  \qw_s^r$, with $\veks \! \triangleq  
\Ex\Big\{ \big\vert \Tilde{\qg}_{sk}^H \Tilde{\qw}_{sk}^{\ZF} - \Ex\{\Tilde{\qg}_{sk}^H \Tilde{\qw}_{sk}^{\ZF} \}\big\vert^2\Big\}$, $\tkis \triangleq \beta_k (\bar{\qw}_{si}^{\ZF})^{H} \bar{\qw}_{si}^{\ZF}$, and $\vrhokjssp \triangleq  \Ex\big\{\tilde{\qg}_{sk}^{H}\Tilde{\qw}_{sj}^{\ZF} (\Tilde{\qw}_{s'j}^{\ZF})^H \tilde{\qg}_{s'k} \big\} $.


For sensing, the transmit beampattern gain, $\qg_{sen}(\bar{\qa},\tilde{\qa})\triangleq \Ex\{\| \qv^T(\theta_s^t, \phi_s^t) \qx_l \|^2\}$ is given as \cite{Xiang:TSP:2020}, \cite{Hua:TWC:2024},
\vspace{-0.7em}
\begin{align}
    &\qg_{sen}(\bar{\qa},\tilde{\qa}, \qa) = \sum\nolimits_{s\in\Set}\qv( \theta_s^t, \phi_s^t)^T \qR_X \qv( \theta_s^t, \phi_s^t)^{\ast} \nonumber \\
    &= \sum_{k\in\Kn}\!\!\Big|\sum_{s\in\Set} \sqrt{\betask}\bas \qv_s^T( \theta_s^t, \phi_s^t) \bwsk^{\ZF} \Big|^2   \!+\! \sum_{s\in\Set} \!\sum_{s'\in\Set}\!\! \qv_s( \theta_s^t, \phi_s^t)^T \nonumber \\
    &\hspace{1em}\Big( \sum\nolimits_{j\in\Kf} \sqrt{\Tetasj \Tetaspj} \tas\taps \Ex\{ \twsj^{\ZF}(\twspj^{\ZF})^H\} \Big) \qv_{s'}( \theta_s^t, \phi_s^t)^{\ast} \nonumber \\
    &\hspace{1em} \! + \! \sum\nolimits_{s\in\Set} \! \eta_s | a_s \qv_s^T( \theta_s^t, \phi_s^t)\qw_s^r |^2,
\end{align}
where $\qx_l = [\qx_{l,1}^T,\ldots,\qx_{l,S}^T]^T\in \mathbb{C}^{M_t \times 1}$ is the transmit signal at the $l$-th time instant and $\qR_x = \Ex\{ \qx_l \qx_l^H \}$ denotes the covariance matrix of the transmit symbols, while $\qv_s( \theta_s^t, \phi_s^t) = [\qv(\theta_s^t, \phi_s^t)]_{\iota_s}$ is the steering vector of the $s$-th subarray. 

The total power consumed by the ELAAs is 
\vspace{-1em}
\begin{align}~\label{eq:PC}
    P_C &= \sum\nolimits_{s\in\Set}\! \frac{1}{\zeta}\gamma_s + 2P_{\text{syn}} + \sum\nolimits_{s\in\Set} a_s M_s P_{\text{ct}},
\end{align}
where $\gamma_s \!\triangleq\! \sum\nolimits_{k\in\Kn}\!\! \bas^2 \bar{\eta}_{sk} \bar{\Psi}_{sk} +  {\sum\nolimits_{j\in\Kf}\!\!\tas^2\Tetasj \Tilde{\Psi}_{sj}} +  a_s^2 \eta_s \| \qw_s^r \|^2$,
$\bar{\Psi}_{sk} = \tr(\bar{\qw}_{sk}(\bar{\qw}_{sk})^H)$, and $\Tilde{\Psi}_{sj} = \Ex \{ \tr(\Tilde{\qw}_{sj}(\Tilde{\qw}_{sj})^H)\}$ \cite{Jun:TCom:2024}; $\zeta$ is the efficiency of the power amplifier, $P_{\text{syn}}$ is the power consumed by the frequency synthesizer and $P_{\text{ct}}$ is the circuit power consumed by the RF chain.

The optimization problem is as follows: 
\vspace{-0.6em}
\begin{subequations}\label{eq:original optimization problem}
\begin{alignat}{2}
(\mathcal{P}1):\hspace{0.2em}&\underset{\{ \bar{\qa}, \tilde{\qa}, \qa \}}{\min}       
&~& P_{C} \label{eq:optProb}\\
&\hspace{0.8em}\text{s.t.}
&         &\sum\nolimits_{s\in\Set} \gamma_s \leq P_t,  \label{eq:overall:power} \\
&&&\gamma_s \leq P_s, ~\forall s\in \Set,  \label{eq:each subarray:power} \\
&         &      &\overline{\text{SINR}}_{k}\geq \bSEkth, ~\forall k\in\Kn, \label{eq:P1:QoS:NF}    \\
&         &      &\widetilde{\text{SINR}}_{k}\geq \tSEkth, ~\forall k\in\Kf, \label{eq:P1:QoS:FF}    \\
&&&\Ex[\| \qv^T(\theta_s^t, \phi_s^t) \qx_l \|^2]  \geq \kappa,  \label{eq:beampattern} \\
&         &      & \bas,\tas \in \{0, 1\},~ \forall s\in \Set,    \\
&         &      & a_s = \min\{1,\bas+\tas\}, ~\forall s\in \Set, \label{eq:a_s requirement} 
\end{alignat}
\end{subequations}
where $P_t$, $P_s$ are the maximum transmit powers of the whole array and each individual subarray, respectively; $\bSEkth$, $\tSEkth$ are the minimum SINR requirements of NFUEs and FFUEs, $\kappa$ is the threshold on the beampattern gain toward the target direction, respectively. The constraint \eqref{eq:beampattern} represents the requirement of sensing performance and \eqref{eq:P1:QoS:NF}, \eqref{eq:P1:QoS:FF} denote the performance requirements of communication users. 

\vspace{-0em}
\section{Proposed Algorithm}
\label{subsec:Optimization}
\vspace{-0.5em}
The problem $({\mathcal{P}1})$ is nonconvex due to the binary variables $\bas, \tas, a_s, \forall s \in \Set$, and the constraints \eqref{eq:P1:QoS:NF}, \eqref{eq:P1:QoS:FF}, \eqref{eq:beampattern} and \eqref{eq:a_s requirement}. To address these, first, we relax the binary constraints to continuous ones, e.g., $\bas, \tas, a_s \in [0,1]$ \cite{Mohammadi:TWC:2025}, and incorporate  three penalty terms as $\lambda_1 \sum\nolimits_{s\in \Set}\bas(1-\bas) + \lambda_2 \sum\nolimits_{s\in \Set}\tas(1-\tas) + \lambda_3 \sum\nolimits_{s\in \Set}a_s(1-a_s)$ into the objective function~\cite{Mohammadi:TWC:2025}. The penalty parameters $\lambda_1$, $\lambda_2$, and $\lambda_3$ balance binary recovery and power minimization. By linearizing the concave penalty terms via first-order Taylor expansion, we obtain the convex objective
\vspace{-0.7em}
\begin{align}
    P_{C_1} &= P_{C} + \lambda_1 \sum\nolimits_{s\in \Set} (1-2 \bas^{(n)}) \bas + (\bas^{(n)})^2 \nonumber \\
    &\hspace{2em} + \lambda_2 \sum\nolimits_{s\in \Set} (1-2 \tas^{(n)}) \tas + (\tas^{(n)})^2\nonumber \\
    &\hspace{2em} + \lambda_3 \sum\nolimits_{s\in \Set} (1-2 a_s ^{(n)}) a_s + (a_s^{(n)})^2,
\end{align}
where we used a superscript ($n$) to
denote the value of the involving variable produced after ($n-1$) iterations ($n \geq 1$).

To deal with non-convex constraint~\eqref{eq:P1:QoS:NF}, we rewrite it as
\vspace{-1.8em}
\begin{align} \label{eq:reformulated NF SINR:SA}
    \betak\big\vert\sum\nolimits_{s\in\Set}\bar{\qg}_{sk}^{H}\bar{\qw}_{sk}^{\ZF}\bas\big\vert^2 \!\geq\! \big( \bar{\Delta}^{\ZF} \!+\! \sigma_k^2 \big) \bSEkth,~\forall k\in\Kn.
\end{align}
We now define $\bar{c}_k^s \triangleq \bar{\qg}_{sk}^{H}\bar{\qw}_{sk}^{\ZF}$, and $\bar{\qc}_k = [\bar{c}_k^s]_{s\in\Set}$. Then, can rewrite the left-hand side of~\eqref{eq:reformulated NF SINR:SA} as  $\bar{f}_k(\bar{\qa}) \triangleq \big\vert\sum\nolimits_{s\in \Set} \bar{c}_k^s \bas \big\vert^2 =  (\bar{\qc}_k^H \bar{\qa}^{\ast})(\bar{\qc}_k^T \bar{\qa})$. To handle the non-concavity of the left-hand side of~\eqref{eq:reformulated NF SINR:SA}, we approximate the convex term $\big\vert\bar{\qc}^T \bar{\qa}\big\vert^2$ by its first-order Taylor expansion at $\bar{\qa}^{(n)}$ as $\bar{f}_k(\bar{\qa}) \geq \bar{f}_k(\bar{\qa}^{(n)}) +  \nabla_{\bar{\qa}} \bar{f}_k(\bar{\qa}^{(n)})^T (\bar{\qa} - \bar{\qa}^{(n)}) \!=\! \big\vert\bar{\qc}_k^T \bar{\qa}^{(n)}\big\vert^2 \!+\! 2\Re\big\{ (\bar{\qc}_k^T \bar{\qa}^{(n)})^{\ast}\bar{\qc}_k^T (\bar{\qa} - \bar{\qa}^{(n)})\big\} \triangleq \bar{f}_k^{\text{lb}}(\bar{\qa},\bar{\qa}^{(n)})$. On the right-hand side of the inequality in~\eqref{eq:reformulated NF SINR:SA}, we notice that the  $\tas \taps$ ($\bas \baps$) terms in the inter-group (intra-group) interference are not convex, when $\text{$s\neq s'$}$. To tackle this issue, we apply the inequality, $ 4xy \! \leq  (x\!+\!y)^2\!-\!2(x^{(n)}\!-\!y^{(n)})(x\!-\!y) \!+\! (x^{(n)}\!-\!y^{(n)})^2$.
\begin{figure*}[t]
\vspace{-0.5cm}
    \centering
    \begin{minipage}[t]{0.32\textwidth}
        \centering
         \includegraphics[trim=0 0cm 0cm 0cm,clip,width=1.1\textwidth]{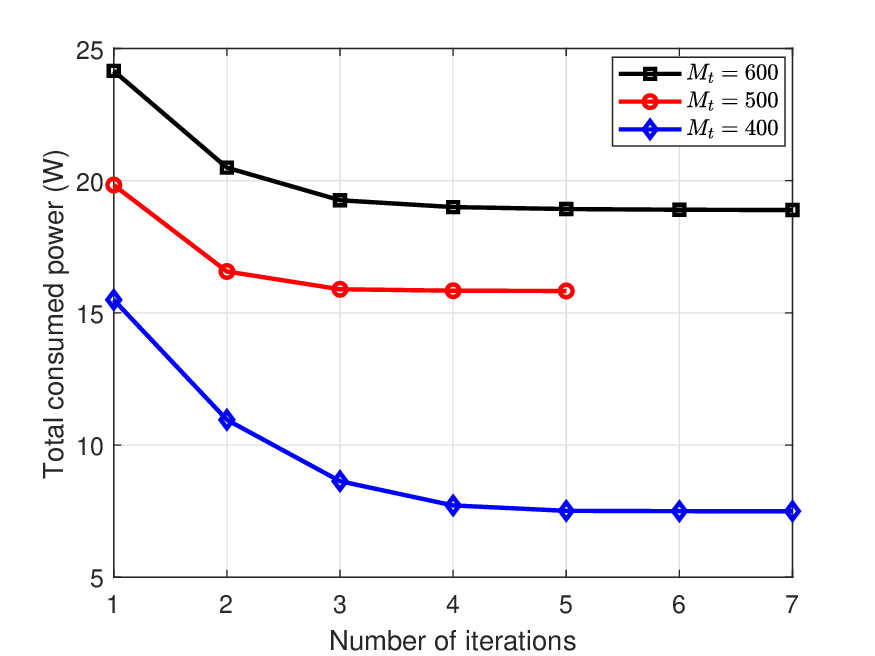}
         \vspace{-1.7em}
        \caption{\small Convergence analysis of the proposed algorithm ($M_s\!=\!50$).} 
        \label{fig:SE_v_thickness}
    \end{minipage}
    \hfill
    \begin{minipage}[t]{0.32\textwidth}
        \centering
          \includegraphics[trim=0 0cm 0cm 0cm,clip,width=1.1\textwidth]{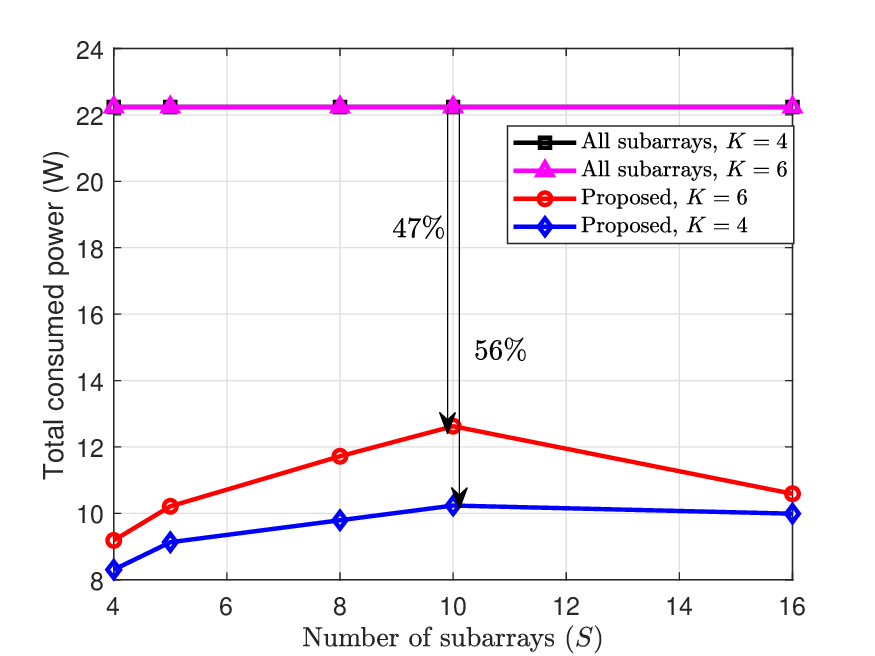}
          \vspace{-1.7em}
        \caption{\small Power consumption for fixed number of antennas ($M_t=400$, $K_F=K_N$).}
        
    \end{minipage}
    \hfill
    \begin{minipage}[t]{0.32\textwidth}
        \centering
         \includegraphics[trim=0 0cm 0cm 0cm,clip,width=1.1\textwidth]{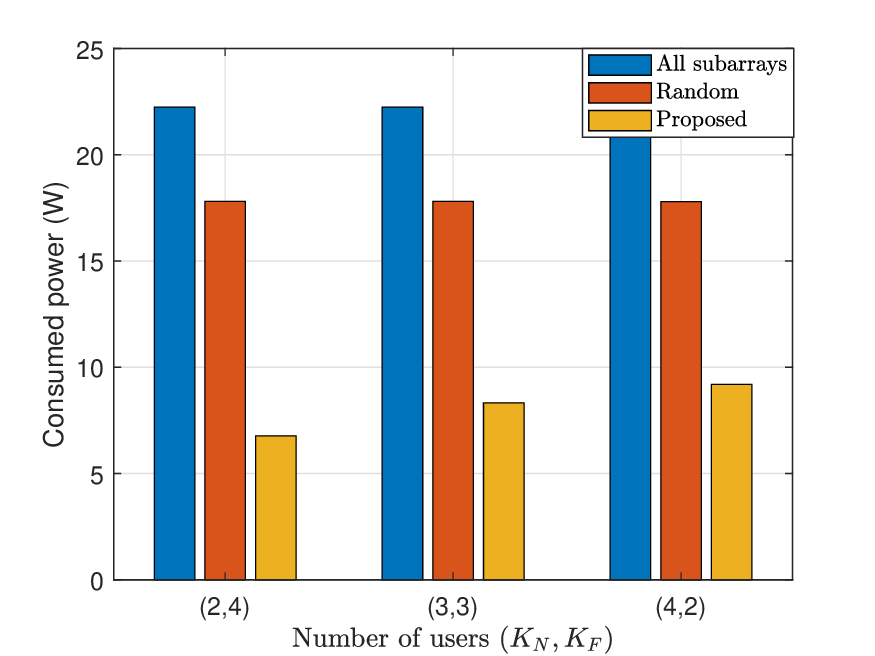}
         \vspace{-1.7em}
        \caption{\small Power consumption with different user configurations ($S$, $M_t=400$).}
        \label{fig:figurec}
    \end{minipage}
\vspace{-1.2em}
\end{figure*}
Using analogous convexification techniques, we transform \eqref{eq:P1:QoS:FF} and \eqref{eq:beampattern} into their respective lower-bound approximations, denoted by $\Tilde{f}k^{\text{lb}}(\tilde{\qa},\tilde{\qa}^{(n)})$ and $\qg_{sen}^{\text{lb}}(\bar{\qa},\bar{\qa}^{(n)},\tilde{\qa},\tilde{\qa}^{(n)}, \qa, \qa^{(n)})$.  Finally, the resulting optimization problem is as follows:
\vspace{-0.5em}
\begin{subequations}\label{eq:updated (P2.3):subarray activation}
\begin{alignat}{2}
(\mathcal{P}2):\hspace{0.2em}&\underset{\{\bar{\qa}, \tilde{\qa}, \qa\}}{\min}       
&~& P_{C_1} \\
&\hspace{1em}\text{s.t.}
&         & \eqref{eq:overall:power}, \eqref{eq:each subarray:power}, \\
&         &      &\betak \bar{f}_k^{\text{lb}}(\bar{\qa},\bar{\qa}^{(n)})\geq \bSEkth\bar{\Pi}_k^{\text{ub}}, \forall k\in\Kn,    \\
&         &      &\Tetak\Tilde{f}_k^{\text{lb}}(\tilde{\qa},\tilde{\qa}^{(n)})\geq \tSEkth\Tilde{\Pi}_k^{\text{ub}}, \forall k\in\Kf,    \\
&&& \qg_{sen}^{\text{lb}}(\bar{\qa},\bar{\qa}^{(n)},\tilde{\qa},\tilde{\qa}^{(n)}, \qa, \qa^{(n)}) \geq \kappa\\
&         &      & 0\leq \bas,\tas, a_s \leq 1, \forall s\in \Set,  \\
&         &      & \hspace{-1em}a_s \geq \bas, a_s \geq \tas, a_s \leq \bas + \tas, \forall s \in \Set, \label{eq: relaxed min} 
\end{alignat}
\end{subequations}
where we relaxed constraint \eqref{eq:a_s requirement} to \eqref{eq: relaxed min} and
\vspace{-0.6em}
\begin{align*}
    &\bar{\Pi}_k^{\text{ub}} \!\triangleq\! 
    \sum\nolimits_{j\in\Kf} \!\! \Tetaj \! \big(\sum\nolimits_{s \in \Set}\! \tas^2\rhokjs \!+\! \tilde{\mu}_{ss'} \rhokjssp \big) + \!\sum\nolimits_{i \in\Kn\setminus k} \!\!\betai \\
    &\hspace{1em}\! \times\Big(\sum\nolimits_{s \in \Set} \bas^2 \vert\bar{\qg}_{sk}^{H} \bar{\qw}_{si}^{\ZF}\vert^2 + \bar{\mu}_{ss'}\Re\{\bgsk^{H}  \bar{\qw}_{si}^{\ZF} (\bar{\qw}_{s'i}^{\ZF})^H \bgspk\}\Big)\!\nonumber\\
    &\hspace{1em}+ \!\sum\nolimits_{s\in\Set}\! \eta_s a_s^2 \vert\bar{\qg}_{sk}^{H} \qw_s^r\vert^2 + \sigma_k^2,
\end{align*}
\vspace{-1.5em}
\begin{align*}
    &\tilde{\Pi}_k^{\text{ub}} \triangleq  
    \Tetak \!\sum\nolimits_{s\in \Set}\! \tas^2 \veks \! + \! \sum\nolimits_{i\in\Kn} \!\! \betai \sum\nolimits_{s\in \Set} \! \bas^2 \tkis \\
    &+\! \sum\nolimits_{j\in\Kf\setminus k}\!\!\!\Tetaj \Big(\sum\nolimits_{s \in \Set} \!\!\tas^2\vrhokjs \!+\! \tilde{\mu}_{ss'} \vrhokjssp \Big) \!\!+\!\! \sum\nolimits_{s\in \Set}\!\! \eta_s a_s^2 \Vert\qw_s^r  \Vert^2, 
\end{align*}
with $\tilde{\mu}_{ss'} = \frac{1}{2}\sum\nolimits_{s \in \Set}\!\!\sum\nolimits_{s' \in \Set, s'>s}\! \big((\tas\!+\!\taps)^2-2(\tas^{(n)}\!-\!\taps^{(n)})(\tas\!-\!\taps) + (\tas^{(n)}\!-\taps^{(n)})^2\big)$ and $\bar{\mu}_{ss'} = \frac{1}{2}\!\sum\nolimits_{s\in\Set} \!\sum\nolimits_{s'\in\Set,s'>s}\\
\big((\bas\!+\!\baps)^2\!-\!2(\bas^{(n)}\!-\!\baps^{(n)})(\bas\!-\!\baps)\!+\! (\bas^{(n)}\!-\!\baps^{(n)})^2\big)$.   

The problem $(\mathcal{P}2)$ is now convex, which can be solved by CVX \cite{cvx} with proposed algorithm shown in \textbf{Algorithm 1}. The problem $(\mathcal{P}2)$ involves $A_v= 3S$ real-valued scalar variables, $A_l= 6S$ linear constraints, $A_q= (K+3)$ quadratic constraints. Therefore, the algorithm for solving problem \eqref{eq:original optimization problem} requires a complexity of $\mathcal{O}(\sqrt{A_l+A_q}(A_v+A_l+A_q)A_v^2)$ in each iteration~\cite{tam16TWC}. In contrast, an exhaustive search incurs exponential complexity, i.e., $O(2^S)$.

\begin{algorithm}[t]
\caption{Proposed SCA-based algorithm for \eqref{eq:original optimization problem}} \label{alg:SCA} 
\begin{algorithmic}[1]
    \State Initialization: set $n=1$, and use all subarrays activated as initial points $(\boldsymbol{\bar{a}}^{(1)}, \boldsymbol{\tilde{a}}^{(1)},\boldsymbol{a}^{(1)})$. Define a tolerance $\epsilon_1=10^{-3}$, the maximum number of iterations $I_1$, and initial penalty value $\lambda_1 = \lambda_2 = \lambda_3 = 10$.
    \State Iteration $n$: solve \eqref{eq:updated (P2.3):subarray activation}.
    Let $(\boldsymbol{\bar{a}}^{\ast}, \boldsymbol{\tilde{a}}^{\ast}, \boldsymbol{a}^{\ast})$ be the solution, and update $\lambda_i=\eta\lambda_i$ for $i\in\{1,2,3\}$, with $\eta\geq 1$.
    \State If the fractional increase of the objective function value is below a threshold  $\epsilon_1 > 0$ or $n=I_1$, stop. Otherwise, go to step 4.
    \State Set $n = n + 1$, update $(\boldsymbol{\bar{a}}^{(n)}, \boldsymbol{\tilde{a}}^{(n)}, \boldsymbol{a}^{(n)}) = (\boldsymbol{\bar{a}}^{\ast}, \boldsymbol{\tilde{a}}^{\ast}, \boldsymbol{a}^{\ast})$, go to step 2.
\end{algorithmic} 
\end{algorithm}
\setlength{\textfloatsep}{0.0cm}

\vspace{-0.6em}
\section{Numerical Results}
\label{sec:simulation}
\vspace{-1em}
We present numerical simulations to evaluate the performance of the proposed algorithm. The large-scale fading coefficient $\beta_k$ is modeled as $10^{-0.53}/d_{k}^{2}$~\cite{Zhang:TCOM:2024}, where $d_k \in [110, 160]$ is the distance from the FFUEs to the array. For a fair comparison, we set $\bSEkth$ and $\tSEkth$ for each FFUE and NFUE to $70\%$ of their respective values when all subarrays are activated. Similarly, $\kappa$ is chosen as $70\%$ of the maximum beampattern gain toward the target direction. We set $M_s = 50$, $\sigma^2 = -104$ dBm, $P_{\text{syn} } = 50$ mW, $P_{\text{CT}} = 48.2$ mW, $P_t=1000$ mW, $P_s=1.5\times\frac{P_t}{S}$ mW, $\zeta = 0.35$, while the penalty factor is $\eta=1.5$ \cite{Zhang:TCOM:2024}. We assume that the target is located in $\theta \in (0, \frac{\pi}{2})$, $\phi \in (-\frac{\pi}{2}, \frac{\pi}{2})$. 
We compare the proposed design against two benchmarks: (i) \textbf{All-subarrays,} where all subarrays are employed for both communication and sensing; and (ii) \textbf{Random,} where the same number of subarrays as in the \textbf{proposed} design are randomly activated. If the constraints are not met, the number of active subarrays is incrementally increased until they are satisfied.

Figure 1 shows the convergence of the proposed algorithm for $M_t = 400$, $500$, and $600$. The total power consumption, i.e., the objective function,  quickly stabilizes, requiring only $7$ iterations for $M_t = 400(600)$ and $5$ for $M_t = 500$, demonstrating the algorithm’s efficiency. As $M_t$ increases, the algorithm converges to a higher stationary point due to more stringent SINR and beam-pattern requirements.

Figure 2 illustrates power consumption versus $S$ and for fixed $M_t = SM_s$. For $S \leq 10$, the total power consumption increases with $S$ due to the reduced beamforming gain of small antenna arrays. As $S$ increases, finer subarray activation enables the algorithm to deactivate more subarrays, thereby reducing overall power consumption. Notably, identical subarrays are selected for NFUEs, FFUEs, and sensing, minimizing the required number of RF chains. 

In Fig. 3, we examine the total power consumption for different user combinations.
The \textbf{Proposed} algorithm significantly reduces power consumption compared to \textbf{All-subarrays} and \textbf{Random} schemes. The consumption increases with more NFUEs due to stricter near-field SINR constraints requiring higher transmission power. 
\vspace{-.3em}
\section{Conclusion}
\label{sec:conclusion}
\vspace{-1em}
We proposed a subarray activation strategy that minimizes total power while satisfying QoS requirements for both communication and sensing in ELAA-assisted ISAC systems. Our results reveal a trade-off between beamforming efficiency and selection flexibility: larger subarrays enhance power efficiency by generating higher gain beams, while more subarrays enable finer activation control. Numerical results confirm the effectiveness of the approach, yielding up to $50\%$ power reduction compared to all-subarrays activation.

\vfill\pagebreak

\bibliographystyle{IEEEtran}
\bibliography{strings,refs}

\end{document}